\documentclass[twocolumn,amsmath,amssymb]{revtex4}

\pdfoutput=1

\usepackage{epsfig}
\usepackage{graphics}

\begin{document}

\title{Measurement of Quantum Fluctuations in  Geometry}

\author{Craig J. Hogan}
\affiliation{University of Washington,  Seattle, WA 98195-1580, USA}

\begin{abstract} 
A particular  form for the quantum  indeterminacy of relative spacetime position of events  is   derived from  the  limits of measurement possible with Planck wavelength radiation.  The indeterminacy predicts fluctuations from a classically defined geometry in the form of  ``holographic noise'' whose spatial character,  absolute normalization, and spectrum  are  predicted  with no parameters.  The noise has a distinctive transverse spatial shear signature, and a flat  power  spectral density  given by the Planck time.  
  An interferometer signal  displays noise due to the uncertainty of relative positions of reflection  events.  The noise corresponds to an accumulation of phase offset with time that mimics a random walk of those optical elements that change the orientation of a wavefront.  It only appears in measurements that compare transverse positions, and  does not appear at all in purely radial position measurements. A lower bound on holographic noise  follows from  a  covariant upper  bound on gravitational entropy.   The predicted  holographic noise spectrum is estimated to be comparable to measured noise in the  currently operating interferometer GEO600. Because of its transverse character, holographic noise is reduced relative to gravitational wave effects in other interferometer designs, such as LIGO, where beam power is much less in the beamsplitter than in the arms. \end{abstract}
\maketitle
\section{Introduction}
There have been to date no experiments revealing  quantum behavior of spacetime.  On the other hand it is possible  that quantum geometrical effects--- new behavior reflecting a quantum indeterminacy of geometry not present in the classical model of  spacetime defined as a smooth manifold with definite paths between pointlike events---    may be directly detectable in the form of spacetime fluctuations, particularly as a new source of noise in interferometers.   Predicted effects, such as noise spectra in various instrumental setups, have been estimated in the context of  a variety of theories of quantum gravity and Lorentz symmetry violation
\cite{ellis,AmelinoCamelia:1998ax,AmelinoCamelia:1999gg,AmelinoCamelia:2001dy,AmelinoCamelia:2003zf,AmelinoCamelia:2005qa,Smolin:2006pa}.   In this paper, a    holographic hypothesis  for quantum geometry   is formulated   that is more specific than previous phenomemological  estimates of quantum-gravitational noise derived from those more general theoretical approaches.   As shown here, this specificity leads   to sharp predictions  for the transverse spatial character,   spectrum and absolute normalization of the noise, and   parameter-free predictions for 
statistical properties of signals in real interferometers.

The motivation for the particular new form of  quantum spacetime indeterminacy considered here
 originates in  ideas for reconciling gravity and quantum mechanics built on the idea of holography\cite{'tHooft:1993gx,Susskind:1994vu,'tHooft:1999bw,Bousso:2002ju,Padmanabhan:2006fn,Padmanabhan:2007en}.   A holographic quantum geometry of spacetime has two spatial dimensions instead of three, and  the apparent third dimension emerges, by a hologram-like encoding, along a null projection of a 2D sheet.
Although holographic noise is not derived here from a fundamental theory, its observable properties are fixed by  simple and general arguments of an essentially geometrical character,  based  on indeterminacy of measurements made using Planck wavelength radiation, with an absolute scale ultimately normalized by covariant bounds on geometrical entropy\cite{Hogan:2007rz,Hogan:2007hc,Hogan:2007ci,Hogan:2007sw}.  Holographic quantum geometry implies a surprisingly large quantum indeterminacy in the relative transverse positions of events.  These appear in the 3D world as  apparent fluctuations  in the metric with     a distinctive, purely transverse  spatial shear character not apparent in predictions of earlier phenomenological analyses \cite{AmelinoCamelia:1998ax,AmelinoCamelia:1999gg,AmelinoCamelia:2001dy,AmelinoCamelia:2003zf,AmelinoCamelia:2005qa},  and a flat,  frequency-independent spectrum, with a  power spectral density of equivalent metric shear fluctuations given simply by the Planck time $t_P$.     

This holographic noise appears to be observable in  the   signals of some operating interferometers.   Although some experiments have been mounted expressly to search for quantum gravitational effects (e.g., \cite{schiller2004}), the most sensitive 
 interferometers are those built to detect gravitational waves\cite{hough}.   These interferometers have now attained an important threshold of sensitivity:  the measured  spectral density of noise in the recent science runs of LIGO\cite{Abbott:2003vs,Abbott:2007tw} is less than $h=10^{-22} {\rm Hz}^{-1/2}$ over a broad band, from about 70 Hz to about 300 Hz.   This number should be compared with
 the square root of the Planck time  $t_P\equiv l_P/c\equiv \sqrt{\hbar G_N/c^5}=5\times 10^{-44}\ \ {\rm Hz}^{-1}$, where $\hbar$ denotes Planck's constant, $G_N$ denotes Newton's constant, and $c$ denotes the speed of light. 
Systems with metric strain noise  below $ h\approx \sqrt{t_P}= 2.3\times 10^{-22} {\rm Hz}^{-1/2}$ in principle  have the capability of ruling out or possibly  studying in detail effects arising at the Planck scale.  The  detectability of such effects is also critically dependent on details of interferometer design: the  holographic noise discussed here
appears only in comparisons of transverse positions,    as occurs for example in reflection off a beamsplitter. For this reason, the most promising currently operating experiment for detecting the effect is not LIGO, but  GEO600  
\cite{GEO}.  The estimate below of the signal in equivalent gravitational wave spectral density predicts that the current GEO600 apparatus  should display measurable holographic noise, and will allow a test of this class of holographic theories.

\section{quantum geometry defined by Planck wavelength radiation}

 The basic physical effect underlying  holographic indeterminacy, uncertainty and noise can  be captured in a simple wave model.
The essential  argument is that  the uncertainty of  spacetime  is determined by the intrinsic indeterminacy of  defining intervals between events using   Planck wavelength radiation.  The only physically significant quantities defining events in a spatially extended spacetime are those that in principle can be mapped using locally measured signals in Planck-wavelength interferometers.  Holographic indeterminacy  arises  as  a consequence of  wave/particle complementarity in a  spacetime defined by such Planck wavelength waves:  it is the limiting precision to which  worldlines  can be measured with Planck radiation.

Consider a metric where the positions of events and paths are defined using only waves longer than a cutoff $\lambda$.  Paths connecting events are then subject to  indeterminacy because of the limitation of defining the endpoints of a ray, or a path, corresponding to any wave. Using the Rayleigh criterion, a particle path corresponding to a wave propagating over a length $L$ has one endpoint   within an aperture of size $D$  and the other within a diffraction spot of size $\lambda L/D$.   The range of possible orientations consistent with those endpoints is minimized when $D=\lambda L/D$.    The endpoints of the ray are uncertain by an amount $\Delta x=\sqrt{\lambda L}$, corresponding to an aperture with the same size as its own diffraction spot at distance $L$. The orientation of a ray of wavelength $\lambda$ over a length $L$ can at best be defined with a precision $\Delta \theta= \sqrt{\lambda/L}$. This criterion defines an unavoidable 
classical transverse indeterminacy of rays that are defined by waves.   We conjecture that the transverse indeterminacy of Planck wavelength quantum paths  corresponds  to quantum indeterminacy of the metric itself.

This argument is classical, but as it is essentially geometrical, it has a wide application. The same indeterminacy generalizes to any theory   limited to quantities that are measured locally by comparing phases in a Planck wavelength interferometer.
It also applies to rays or paths in a virtual 3D world encoded in a 2D hologram. It seems likely to be a feature of 3+1D spacetime emerging as a dual of quantum theory on a 2+1D null surface. As shown below,  less uncertain  transverse position would imply a number of degrees of freedom, as measured by the number of  distinguishable position eigenstates,  in excess of covariant holographic entropy bounds.

Consider events corresponding to interactions of Planck radiation on a null   surface at two different times, at normal coordinates $z_1$ and $z_2$ in a particular frame.  The radial separation $z_1-z_2$  is measured to Planck precision, defining the relative  $z$ position between the events on each surface.     Consider an event on surface $1$,  such as a Planck particle reflection.  The  particle  obeys the usual  Heisenberg commutation relation between conjugate momentum and position observable operators along the transverse $x$-axis:
\begin{equation}\label{heisenberg}
[\hat x(z_1),\hat p_x(z_1)]=-i\hbar.
\end{equation}
The transverse momentum  $p_{x}(z_1)$ of the particle on surface 1 is related to a transverse position displacement on surface 2 by the angular deflection,
\begin{equation}\label{deflection}
 p_{x}(z_1)l_P/\hbar= x(z_2)/ (z_2-z_1).
\end{equation}
Combining equations (\ref{heisenberg}) and (\ref{deflection}) 
yields a commutation relation between transverse position operators for events on the two surfaces, fixed by an elapsed time in the $z$ direction,
\begin{equation}\label{noncommute}
[\hat x(z_1),\hat x(z_2)]=-i l_P (z_2-z_1),
\end{equation}
where Planck's constant $\hbar$ has dropped out.

This formula specifies the complementarity of the transverse position observables at macroscopic separaration along a null trajectory,   and thereby expresses the uncertainty of the null trajectory itself;  quantum indeterminacy prevents more precise specification of the relative transverse positions of events. An interferometer with a beamsplitter can fold the beam and compare the two surfaces and the relative positions of events, $ x(z_1)-x(z_2)$, directly; but even a Planck interferometer cannot eliminate the indeterminacy.

In the usual way, the indeterminacy (Eq. \ref{noncommute}) yields a Heisenberg uncertainty relation:
\begin{equation}\label{uncertainty}
\Delta x(z_1)\Delta x(z_2)>l_P (z_2-z_1)/2,
\end{equation}
where $ \Delta x(z_1),\Delta x(z_2)$ denote the standard deviations on surfaces 1 and 2 of the wavefunction of transverse position measurements.  The standard deviation $\Delta x_\perp$ of the  difference in relative transverse positions   is then given by $\Delta x_\perp^2=  \Delta x^2(z_1)+\Delta x^2(z_2)$; it has a minimum value when $ \Delta x(z_1)=\Delta x(z_2)$.
This defines a  ``holographic uncertainty principle'' for relative  transverse positions   at events of null spacetime separation  and spatial separation $L$ in a given frame:
\begin{equation}\label{uncertain}
\Delta x_\perp^2>l_PL.
\end{equation}
From this we also derive a minimum uncertainty in angular orientation of a null ray of length $L$ along  each transverse axis:
\begin{equation}\label{angledelta}
\Delta \theta_x> \sqrt{l_P/L}.
\end{equation}

 This geometrical wave model quantitatively reproduces the spatial character of holographic indeterminacy without explicitly introducing a holographic principle or even referring to gravity directly, aside from the introduction of the Planck scale as a fundamental wavelength.  Although the quantum evolution of such a system is deterministic and unitary, in the usual quantum mechanical way, measurement choices determine which branch of a wavefunction an observer lies on and break the symmetry of the wavefunction\cite{zurek}.  The choice of orientation of a mirror at a given event--- which determines the direction in which an apparatus measures that event's spacelike position relative to future events--- fixes an eigenstate and collapses the future metric into states compatible with that measurement.  In  orthogonal directions the state is a superposition so measurements of an orthogonal position are indeterminate.  
 
 This quantum behavior differs significantly from the classical background metric assumed in quantum field theory.
Angular uncertainty increases with smaller $L$, so that  a classical spatial direction is   ill defined at the Planck scale and only becomes well defined after many Planck lengths of propagation.  
Angles become better defined at larger macroscopic separation--- in this sense the world becomes ``more classical'' as it becomes on larger scales  ``more three dimensional.''  What is surprising  is that  transverse positions in terms of absolute length actually  become less well defined at larger separations.   Transverse positions of  macroscopically separated events      do not exist as separately observable quantities, but are complementary: knowledge of one position precludes accurate knowledge of the other.     Macroscopic spacetime limited by Planck scale measurements   exhibits  quantum departures from classical Euclidean behavior  on scales significantly larger than the Planck length.

Measurements made locally on a single, spacelike slice of a null surface can be made at Planck resolution and collapse the system into a corresponding eigenstate.  Events on the null projection of the surface--- in the third spatial dimension--- are encoded in this system at the same time.  But in a particular eigenstate those remote events are fuzzy; they are defined with substantially worse than Planck resolution. This notion of a transverse position uncertainty  of events, described as  wavefunctions of interferometer optical surfaces, substitutes here for full correlators of observables.

Although we have treated this effect as a quantum indeterminacy, the same phenomenology can  also be described as a Brownian-like random shear of null surfaces: { Null surfaces (and null particle paths) execute a Planck random walk in the direction  transverse to their direction of propagation}.  The displacement in the transverse direction
 accumulates over a path like a diffusion process; it does not cancel or converge to a classical value, which in any case could not be defined or measured to better precision even with Planck wavelength radiation.  (The angular direction of a path does however converge to a well defined classical value on large scales, which is why a 3+1D approximation works well.) Similar stochastic descriptions  are familiar in quantum systems\cite{nelson66}.
 
\section{Holographic indeterminacy of signals in an idealized interferometer}

\subsection{Quantum Indeterminacy of Reflection Event Positions}

The interferometer model provides a concrete framework for discussing  this particular hypothesis about the quantum uncertainty of spacetime, and predicting observable effects. 
The model is built on the idea of a machine   where the optical elements share the same transverse position uncertainty as Planck quanta,   an  indeterminacy that reflects   quantum indeterminacy of spacetime itself.  Relative positions of events are only physically distinguished  to the extent  that they can be measured with the  Planck interferometer.
 This notion of spacetime position implies  substantially greater  indeterminacy in the transverse positions of events than  the definition of intervals using null trajectories  to connect worldlines and local clocks to measure separation of events\cite{wigner,salecker}, or the limits imposed locally by measurements with Planck radiation\cite{padmanabhan87}.  The underlying reason for the noise is   indeterminacy of  3D states defined by  data on a two-dimensional surface, a property also illustrated  by the holographic character  of states in a field theory with a UV cutoff\cite{Srednicki:1993im,Yarom:2004vp}.

In practical terms, measurement of holographic noise requires a comparison of transverse positions over a  macroscopic interval, using elements of an interferometer--- a real one, not one using Planck radiation.  Interferometer phase measures the  relative positions of events defined by  photon interactions with  pieces of macroscopically separated  optical elements.    The effect of the holographic noise is to add indeterminacy to the measurements of the position differences in   two orthogonal directions.  The hypothesis is that a real interferometer cannot have less noise than one using Planck radiation.

\begin{figure}
\epsfysize=2in 
\epsfbox{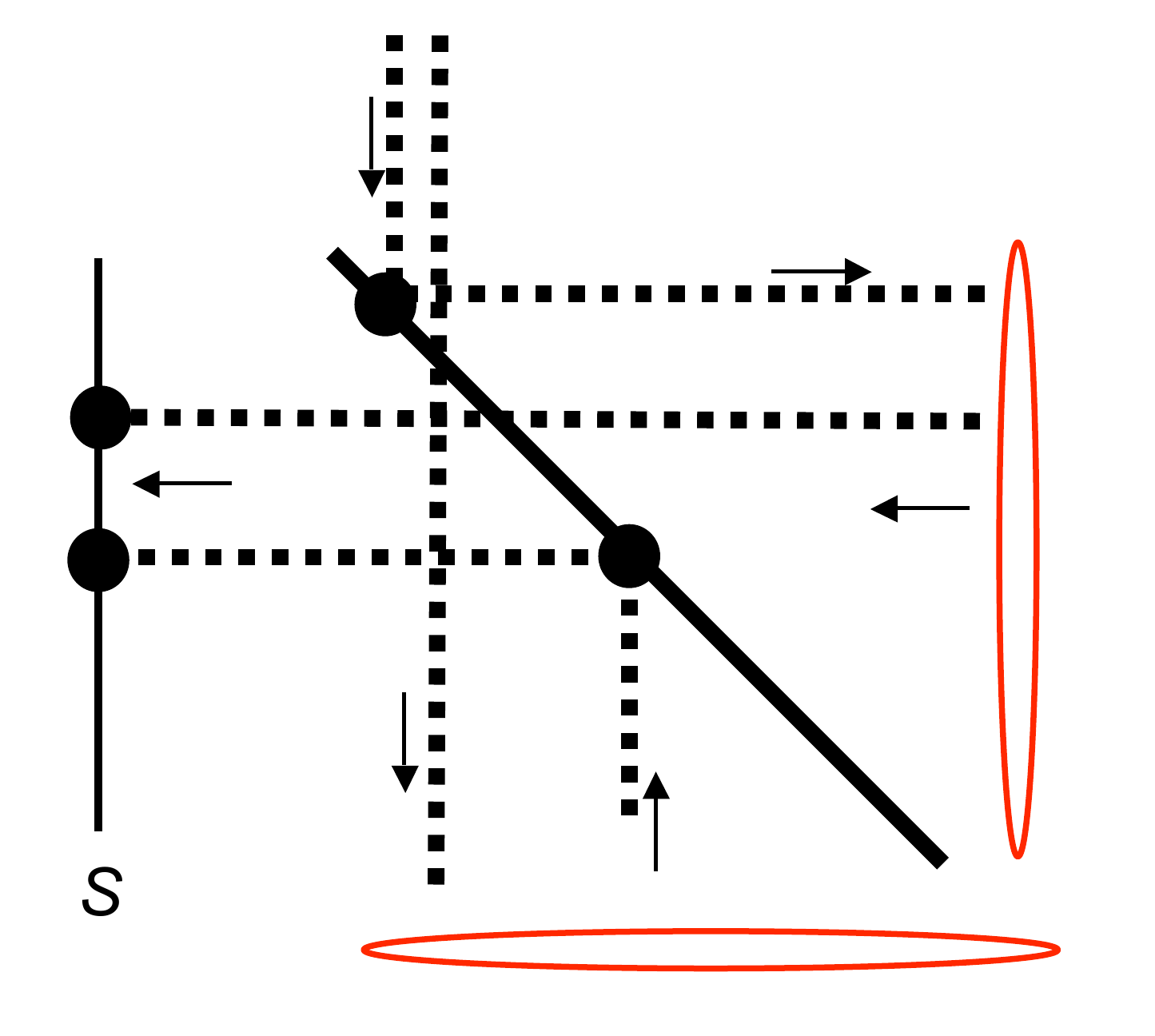} 
\caption{ \label{path}
Inteferometer showing two branches of a single self-interfering particle path with  dotted lines. Reflection events on the beamsplitter, and detection events on the detector $S$, are shown by dots.   After the  return reflection, the incoming paths have an indeterminate transverse position indicated by the elliptical contours.}
\end{figure}

Consider  a quantum-limited Planck wavelength interferometer with a beamsplitter (Figure \ref{path}).     A   quantum enters the apparatus at the top, and eventually is detected on the surface $S$.      Its path is split into beams travelling down and right, sent  in those directions and then returned after travelling a distance $L$ out and back. After   recombining  at the beamsplitter, its arrival is recorded on a detector surface $S$.  Each interfering branch of the particle path involves one reflection event on the beamsplitter,  and the transverse location of the particle is indeterminate until it is detected. Classically, the arrival time of the particle does not depend on its transverse position: any of the various paths from one of the distant directions leads to an arrival at the same time. An interferometer measures the variations in the path difference to the two distant reflections by an average signal strength on $S$; for a single particle the arrival time probability  distribution is proportional to signal strength.

The particle wavepacket  is broadened into ellipses as shown in 
    
(\ref{path}), but  remains narrow in the propagation direction so the  arrival at $S$ happens ``at the same time'' (in the lab frame) whichever path is taken, even though   the transverse arrival position is  uncertain by  $\Delta x=\sqrt{l_P L}$.   Arrival events for various paths of a particle  in   a given  wavefront have a spacelike separation and the invariant interval between them is distributed with a variance 
$\Delta x^2 \approx {l_P L}$.  Classically,   detection events  on $S$   inhabiting a   null surface   identified with a particular wavefront  have  spacelike separation from each other but all happen at the same time in the lab frame. The beamsplitter simply changes the propagation direction of the wavefront without changing the   separations of the events normal to the surface.

 In a holographic theory  the reflection and detection events have a different  relationship since in  the holographic world everything lives on null surfaces, including the inclined reflecting surface of the beamsplitter. For a world description based on a null surface, the reflection events at different  places on the 2D reflecting surface inclined with respect to the null surface do not behave with respect to each other like classical pointlike events.    While the relative positions of   events   on one slice of the 2D null surface (corresponding to  a wavefront at one time in the lab frame)   have relative transverse locations defined to Planck precision, the transverse positions of other events on different slices of the same surface, projected into the third (virtual) dimension, are  not pointlike but are indeterminate.    Their 3D location is encoded nonlocally in a patch or wavepacket of a size that grows with the separation between slices.  In 3D space the patches correspond to wavefunctions that encode relative transverse positions of events, such as those of the same wavefront reflecting off different parts of the beamsplitter, located  on different slices of the null surface.    (In holographic dual models this indeterminacy is described using using quantum correlators.)

  Holographic  uncertainty shares much in common with the fuzziness seen in real holograms.  A simple  description of  holographic uncertainty  is that transverse spacetime positions themselves share  the same indeterminacy that would attach to  Planck wavelength quanta.  Thus the position of an optical surface is uncertain between outgoing and incoming reflections by $\Delta x=\sqrt{l_P L}$ where $L$ is the path length between reflections.  A transverse uncertainty of beamsplitter position relative to either arm  gives the same effect on the final signal:  measured arrival times of wavefronts separated by $t$ have
an uncertainty   with standard deviation $\Delta t \approx \sqrt{t_P t}$. 

\subsection{Random Walk of Signal Phase}

   The position of any body, including an optical element such as the beamsplitter  of an interferometer, is encoded holographically on a null surface. If we choose the surface to coincide with incoming quanta, the reflection events are in eigenstates of position in that direction.  The quanta are then reflected in an orthogonal direction, placing the reflection event in an eigenstate of position in an orthogonal direction.
 Each effectively independent measurement of a quantum arriving on surface $S$   resets the metric  to a new eigenvalue of relative  position between $S$ and a beamsplitter element.   Each independent reflection event adds an uncertain arrival time offset and these offsets accumulate  in quadrature, instead of  averaging  to a fixed classical value as they would in a spacetime with a classically fixed beamsplitter position.
  The beamsplitter position as  determined by a phase signal measuring a difference in position in the two directions appears to undergo a random walk with time.  
   
\begin{figure}
\epsfysize=2in 
\epsfbox{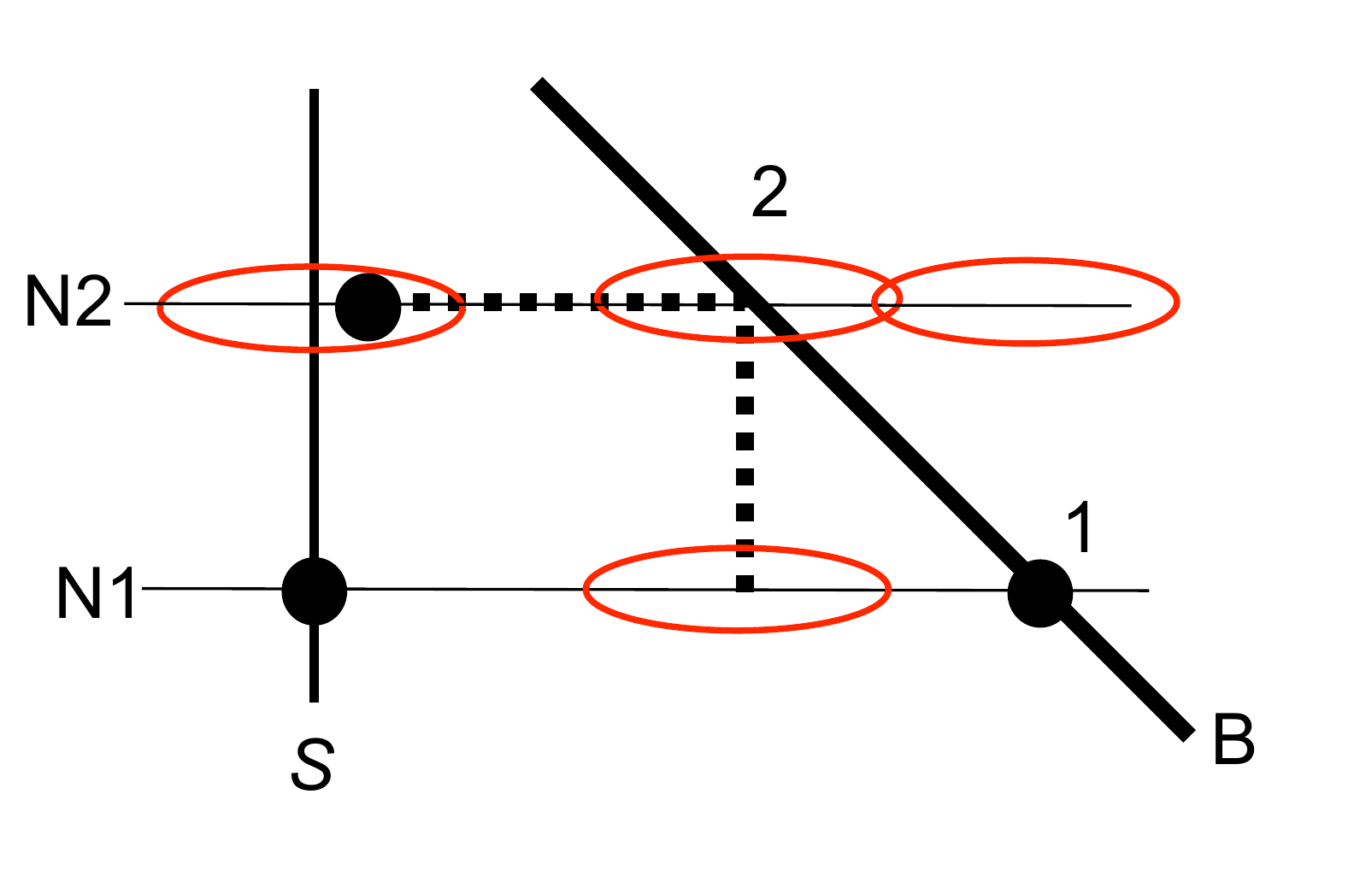} 
\caption{ \label{null}
Holographic encoding of two reflection events on the beamsplitter B. A null surface N is shown at   slice N1 corresponding to reflection event 1 and slice N2 corresponding to event 2.  Once the position of event 1 is fixed, the relative position of a   reflection event 2 on a different slice of the same null surface (the same wavefront in 3D) is encoded by a patch of fields with a finite transverse size given by holographic uncertainty, indicated by an ellipse.  The reflection   rotates the wavefront to land on a detector surface S where measurement collapses the uncertainty of event location in that direction to a definite value.   Again fixed to a reference time from event 1,  the arrival time from event 2  inherits the holographic uncertainty in transverse position,   observed as a real difference in arrival times from the two events.}
\end{figure}

 When we compare the descriptions of the same world on two orthogonal null surfaces, the relative positions of events   specified to lie on a single slice of one null  surface are  on different slices of the other.    When a reflection occurs, a system in an eigenstate of one direction is measured (and ``collapses'') into an eigenstate  in another direction.     An  uncertainty of  relative position in the original null surface coming from below (say)    converts to a spread of arrival times in the orthogonal direction (see Figure \ref{null}).
The  uncertainty of the relative event positions changes depending on a measurement choice of null surface orientation.

Let $\theta$ denote the angle of incidence relative to normal.
Suppose  that every  sampling time $t_*$ in the incoming wave an independent error adds holographic  uncertainty in the orthogonal direction given by $\Delta t_i^2={t_*t_P}$, as discussed above (Eq. \ref{uncertain}).
On the reflecting surface, the separation between these events is
$t_*/\sin\theta$.  
 The number of terms $N$  in the summed phase offset between measured wavefronts 
   is the number of independent events between them. 
 Between two wavefronts separated by time $t$   we 
  then have $N= t\sin\theta/t_*$ offsets added in quadrature.
The standard deviation of position indeterminacy from each independent event orthogonal to the incoming wave is $\Delta t_i$;  when  projected into arrival time in the direction of the outgoing reflected wave, the standard deviation in arrival time from each one is $\Delta t_i\sin2\theta$.  
 The  wavefront arrival time after time $t$ thus differs   by a gaussian random offset from the previous wavefront, with a standard deviation
 \begin{equation}\label{noise}
 \Delta t=\sqrt{ t_Pt}\sqrt{\sin\theta}\sin2\theta.
 \end{equation}
 The actual sampling time $t_*$ cancels out in this result. The interpretation of Eq.(\ref{noise}) is that a
    signal from beam  interference     wanders randomly:   statistical fluctuations    mimic the effect of a random walk of the  beamsplitter surface of  about a Planck length per Planck time, times some geometrical factors.    Note that in this model and in these units, the  noise in the signal depends only on the geometric layout of the optical elements, and not on any other parameters such as power, arm length or wavelength.

 The geometrical factors in Eq.(\ref{noise}) show how the effect varies for different kinds of reflections.  The noise does not appear  in  normal-incidence   optical cavities of an interferometer,  and is highly suppressed in near-normal or grazing-incidence reflection.     In  these cases the transverse position is still uncertain, but it is not measured in the arrival time signal.   The geometrical factors in (\ref{noise}) affect the absolute normalization of holographic noise and quantitative tests of holographic theories with interferometer data: the full holographic indeterminacy is not detected, but is always suppressed by  a numerical factor depending on the setup.
 
 \subsection{Lower Bound on Holographic Noise from Upper Bound on Gravitational Entropy}
 
  The holographic  hypothesis is that the world can be described by a quantum theory of energy and spacetime   on 2D null surfaces with one quantum degree of freedom per $4l_P^2$, 
where $t_P\equiv l_P/c\equiv \sqrt{\hbar G_N/c^5}$. We refer to the literature for   reviews\cite{Bousso:2002ju,Padmanabhan:2007en} of the  circumstantial theoretical evidence for this idea  from black hole thermodynamics and other arguments.

The normalization of Eq.(\ref{noise})  has been chosen to provide a   lower bound       on   noise predicted by theories  where the effective quantized states of the system in 2D that specify the 3D geometry take the form of  waves or fields in the conventional sense, with a UV cutoff that limits the number of degrees of freedom to that required to satisfy holographic entropy bounds.   
 If  covariant entropy bounds are explained in terms of  a limited number density of degrees of freedom in a theory built on  light sheets\cite{Bousso:2002ju}, arguments based on black hole physics suggest that the number of degrees of freedom must be  less than one per $4l_P^2$, or one per $2l_P$ in each dimension.  For a theory   with a minimum cutoff wavelength $\lambda_{min}$, the Nyquist-Shannon sampling criterion implies that any state  is specified by  its value at two points  in position space  per $\lambda_{min}$.    
 The  normalization of the noise  adopted  in Eq.(\ref{noise}) assumes  a maximum of one independent transverse position measurement   per  $l_P$ in each dimension on the light sheet, with the sampling interval set to the Planck time $t=t_P$.
That is sufficient to specify  states in a theory based on waves with a cutoff at $\lambda_{min}= 2l_P$: it is  thus
the smallest noise compatible with having  one degree of freedom per $\lambda_{min}=2l_P$ in each transverse direction, in a theory where a wave mode corresponds to a quantized degree of freedom.

An experimental upper bound on holographic noise   gives a lower bound  on  the  number density  of independent position eigenstates on a null surface, therefore on  the number of degrees of freedom.   
The noise in Eq.(\ref{noise}) can therefore be regarded as a  prediction of the hypothesis that the  holographic behavior of General Relativity is explained in terms of the quantum behavior of fields on 2+1D light sheets, where the covariant entropy bound arises from a minimum wavelength.

 \section{Predicted Holographic noise in interferometer signals}
 
 \subsection{Michelson Interferometer}

The currently operating interferometer   GEO600 continuously measures the  difference of proof mass positions in   two orthogonal directions from a beamsplitter, where all components are suspended and effectively in free fall in the measured directions\cite{hough,GEO}.
The   holographic uncertainty   is   predicted to appear as a noise in the phase signal stream resembling the effect of a classical random walk of the beamsplitter position, along its inclined axis mixing the two orthogonal directions.

The accumulated phase difference between the $AB$ and $BC$  arms of a simple Michelson interferometer is the same as it would be if there were a classical random perturbation of the difference in arm lengths, $\Delta l= AB-BC$, at time $t$, with variance
 \begin{equation}\label{predict}
\langle \Delta l^2\rangle_H >c t l_P {\sin\theta}(\sin2\theta)^2.
\end{equation}
The zero-parameter prediction (Eq. \ref{predict}) for holographic noise in terms of equivalent beamsplitter position noise, in absolute physical units, is the main result of this paper.  It provides a precise    experimental target for  direct test of the holographic hypothesis for Planck scale quantum geometry represented by Eq.(\ref{noncommute}).  Should holographic noise exist,  its universal spectrum, and the specific dependence on transverse measurement characteristic of  its origin,  allow diagnostic signatures that distinguish and separate it from other sources of system noise.  

\subsection{Equivalent Metric Strain Noise Spectrum for Signal-Recycled Michelson Interferometer}

The prediction (Eq. \ref{predict}) for  extra noise in the orthogonal path difference can  be  translated into an equivalent gravitational-wave strain spectrum to compare with  conventional units for quoted noise.  A precise calculation requires a transfer function derived from a detailed model of signal formation in the apparatus. We offer a rough estimate here for a simplified model of the GEO600 signal-recycled Michelson interferometer to show that the prediction is a realistic experimental target.

Consider first  a frequency $f=c/L$,  where  $L$ denotes the interferometer arm length.  
We characterize the holographic fluctuations by $h_H^2$, the equivalent gravitational-wave metric strain perturbation power spectral density per bandwidth--- the gravitational wave spectrum that would give rise to the same signal fluctuations. Consider the passage of a  gaussian gravitational wavepacket of frequency $\approx f$ and duration $t\approx 1/f$, normal to the interferometer plane.   By the standard definition of $h$, the passage of this  gravitational wavepacket  over about one wave cycle  causes a variation of fractional path difference between arms  related to the broadband spectral density $h^2$ by 
\begin{equation}\label{strain}
\langle \Delta l^2\rangle_{GW}/l^2|_{f=1/t}\approx h^2 f,
\end{equation}
where $\langle \Delta l^2\rangle_{GW}$ is the variance in path difference.  Equating this with 
the holographic uncertainty  in beamsplitter position (Eq. \ref{predict}) for $t=L/c$,  we find that the power spectral density per bandwidth of holographic noise at frequency $1/L$ is  given by the Planck time, $h_H^2\approx {t_P}$.  

For lower frequencies, corresponding to many reflections of light in the GEO cavity,  the phase signal in GEO   approximately represents the average arm length difference measured over  time $\approx f^{-1}$.  Light arriving at the detector at a given time combines the phase offsets from many traversals of the arms.  The position  of the beamsplitter at a given instant contributes to--- is measured by---  this instantaneous phase signal not just once, but many ($\approx c/fL$) times, and it has the same value each time. Holographic uncertainty applies only to one measurement of  given event position.     In the combined signal,  the effect of averaging over many reflections is thus to reduce the total phase offset instead of adding offset in quadrature.  
Instead of displaying quantum indeterminacy corresponding to the time  $\approx f^{-1}$, the   difference in phase offset between two times separated by less than a typical photon residence time  tends to converge to a mean ``true'' classical value of the arm difference.    The arm length difference standard deviation at frequency $f=1/t$, with averaging time $t$,  is then $\Delta l\approx (L ct_P)^{1/2} (L/ct)^{1/2}$.  (The measured arm length difference for frequency $f$ is a fraction $fL/c$ of what it would be if the arm length were $c/f$, the same as for a gravitational wave of the same frequency.)   Recalling that the equivalent gravitational wave power spectral density $h$ gives $(\Delta l/l)^2\approx h^2{f}$,   we find  that  the equivalent spectrum remains flat,  $h_H^2\approx {t_P}$, independent of $f$.

At still lower   frequencies, below the   inverse residence or averaging time $t_{res}=1/f_{res}$ contributing to  the location of a fringe, the effective spectrum changes slope.  Phase offsets at widely separated times are   independent measurements.   Holographically generated phase error   accumulates like a random walk with time, leading to $\Delta l\propto f^{-1/2}$ and  $h_H\propto 1/f$.  
The overall holographic noise spectrum is thus given by
\begin{equation}
h_H \approx \sqrt{t_P}
\end{equation}
 for $f> f_{res}$ and
\begin{equation}
h_H \approx (f_{res}/f)\sqrt{t_P}
\end{equation}
for $f< f_{res}$.  
  For GEO600, $f_{res}$ can be estimated from the bandwidth of the signal recycling cavity,  about 700 Hz.
This rough estimate omits numerical factors of the order unity for both $f_{res}$ and overall amplitude, but contains no truly free parameters.   Since it predicts noise   comparable in magnitude and spectrum to the level of  noise measured in the current GEO600 system from about 100 to 600 Hz\cite{strain},  we conclude that   this machine is likely to be capable of either ruling  out  holographic complementarity (Eq. \ref{noncommute}) as a model for quantum geometry,  or of studying quantum geometrical fluctuations in detail.  The approximate  agreement of predicted holographic noise with otherwise unexplained noise in GEO600 motivates further study.

\subsection{Other Interferometer Designs} 

In other interferometer designs   the lengths of two arms are measured separately but are not continuously compared using a beamsplitter.   The difference of independently measured arm lengths  is sensitive to the classical metric  distortion caused by  gravitational waves but not to quantum  holographic noise. Holographic noise is only introduced when orthogonal positions are compared;  except for the occasions where reflection off a beamsplitter    contributes a transverse position uncertainty to phase, holographic noise is not added to the phase signal. 

For example,  the  current LIGO design\cite{Abbott:2003vs,Abbott:2007tw} tends to suppress holographic noise relative to gravitational wave signals.    LIGO's power is much greater  in its two separate Fabry-Perot arm cavities than at the beamsplitter, whereas GEO600  sends the full interferometer power through the beamsplitter.    In LIGO, the gravitational wave signal, but not the holographic noise,  is magnified  by the many normal-incidence bounces the light makes in the cavities.  For this reason, even though LIGO's noise level for gravitational wave detection is already well below an equivalent metric strain noise $\sqrt{t_P}$,  it does not rule out universal holographic noise at its current sensitivity.  

There should similarly be  no observable holographic noise in the baseline design for the proposed space mission LISA in Michelson (two arm) mode, since it compares phase signals from separate  cavities, each of which has only normal incidence reflections. Further analysis is required to estimate the effect for LISA in three-arm Sagnac mode, which compares phases from travel in two directions around a closed triangle. In any case the holographic noise at LISA frequencies is likely to be significantly below the confusion noise generated by real  cosmic gravitational waves.

If the effect is detected, a new interferometer could be designed specifically to measure the properties of holographic noise. Unlike  gravitational waves, measurement of holographic noise does not require a long baseline interferometer.  A system built for the purpose of  studying holographic noise can have  shorter arms than gravitational wave detectors, and therefore use  a smaller vacuum system.  Holographic noise can   be studied above a kilohertz,  where natural sources of classical gravitational waves become extremely weak and there has been little motivation to develop gravitational wave detectors.  However, high frequency studies will be limited by photon shot noise, and require laser power even larger than current interferometers.

\acknowledgements
The author is grateful for conversations and suggestions from  M. Cerdonio, K. Danzmann, S. Hild,  H. L\"uck, D. L\"ust, T. Padmanabhan,  B. Schutz, and S. Vitale; for hospitality of 
the Max-Planck-Institut f\"ur Astrophysik, Garching, and the Albert Einstein Institute, Potsdam; and for support from the Alexander von Humboldt Foundation.

\end{document}